# PropTech for Proactive Pricing of Houses in Classified Advertisements in the Indian Real Estate Market

Sayan Putatunda

Member, IEEE

## Abstract

Property Technology (PropTech) is the next big thing that is going to disrupt the real estate market. Nowadays, we see applications of Machine Learning (ML) and Artificial Intelligence (AI) in almost all the domains but for a long time the real estate industry was quite slow in adopting data science and machine learning for problem solving and improving their processes. However, things are changing quite fast as we see a lot of adoption of AI and ML in the US and European real estate markets. But the Indian real estate market has to catch-up a lot. This paper proposes a machine learning approach for solving the house price prediction problem in the classified advertisements. This study focuses on the Indian real estate market. We apply advanced machine learning algorithms such as Random forest, Gradient boosting and Artificial neural networks on a real world dataset and compare the performance of these methods. We find that the Random forest method is the best performer in terms of prediction accuracy.

## 1. Introduction

Property Technology (PropTech) is the next big thing to disrupt the real estate market. PropTech espouses the use of technology to facilitate the management and operations of real estate assets [1]. The assets here mean either buildings or cities. PropTech can be seen as a part of digital transformation in the real estate industry and it focuses on both the technology and mentality changes of the people involved (including the customers) in this industry [2]. PropTech can also lead to new functionalities such as more transparency, which were not possible earlier [3]. Big data, data analytics, machine learning, blockchain and sensors form a part of PropTech as well [3].

Most of the applications of PropTech documented in the literature are mainly in construction and real estate. One interesting use case is the Rental platforms and "space as a service" that matched supply and demand and work towards the demand for flexibility [3]. Other applications include Sales Platforms and webshops, digitization of administrative management and smart buildings [3]. Please refer to [3] for a detailed overview of various applications of PropTech.

We already see a lot of investments flowing in the PropTech startups in US and European real estate markets [3]. Some of the most innovative PropTech firms across the globe are Atlant, Bowery, Buildium, Flip, Foyr, hOM, Huthunt, No Agent, Opendoor, Open Listings, PeerStreet, Purplebricks, Ravti, Reposit, RealtyShares, Riley, Squarefoot and VR Listing [4]. As far as India is concerned, there is still lot of catch up to be done. One interesting startup that is using PropTech in the Indian context is "Gharvalue". According to their website [5], they have built India's first user driven Automated Valuation Model (UAVM), which can help in performing realistic valuation of a property that can help the customers. As mentioned in their website, "GharValue's UAVM will deliver an online housing value estimate based on proprietary models built on housing data, economic data, neighborhood data, amenities, parking, distances from transportation hubs, malls, schools and myriad other variables." [5]. There are some companies such as unissu.com that are connecting property with PropTech industries [6].

---

*\*This study is a proof-of-concept (PoC) and can be treated as a Technical Report. The purpose is to raise awareness about PropTech by solving one use-case.*

In this paper, we will work on the problem of proactive pricing of houses listed in classified website advertisements in the Indian real estate market as explained in details in Section 2. The rest of this paper is structured as follows. In the next section we will discuss the problem statement and a brief overview of the literature. Then in Section 3 we will discuss the dataset, which is followed by the Experimental results in the next section. Finally, Section 5 concludes the paper.

## 2. Problem Statement & Related Work

In India, there are multiple real estate classified websites where properties are listed for sell/buy/rent purposes such as 99acres [7], housing [8], commonfloor [9], magicbricks [10] and more. However, in each of these websites we can see lot of inconsistencies in terms of pricing of an apartment and there are some cases when similar apartments are priced differently and thus there is lot of in-transparency. Sometimes the consumers may feel the pricing is not justified for a particular listed apartment but there no way to confirm that either. Proper and justified prices of properties can bring in a lot of transparency and trust back to the real estate industry, which is very important as for most consumers especially in India the transaction prices are quite high and addressing this issue will help both the customers and the real estate industry in the long run. We propose to use machine learning and artificial intelligence techniques to develop an algorithm that can predict housing prices based on certain input features.

The business application of this algorithm is that classified websites can directly use this algorithm to predict prices of new properties that are going to be listed by taking some input variables and predicting the correct and justified price i.e. avoid taking price inputs from customers and thus not letting any error creeping in the system. This study on proactive pricing of houses in the Indian context has never been reported earlier in the literature to the best of our knowledge.

However, the problem of house price prediction is quite old and there have been many studies and competitions addressing the same including the classic Boston housing price challenge on Kaggle [11]. As far as housing price prediction in Indian context is concerned, [12] used machine learning techniques such as XGBoost for predicting housing prices in Bengaluru, India. MachineHack [13] conducted an hackathon on predicting housing prices in Bengaluru in 2018. The problem statement was to predict the price of houses in Bengaluru given 9 features such as area type, availability, location, price, size, society, total square foot, number of bathrooms and bedrooms. In fact, the experimental design of this study is motivated by [12] and [13]. Moreover, there have been other studies for house price prediction in other cities of India such as Mumbai [14].

## 3. Data

When someone visits a classified website such as commonfloor [9] to check apartments listed for sell, some of the information that are displayed are the Project name, region, type of apartment (2BHK, 3BHK and more), super built up area of the apartment, carpet area, number of bedrooms & bathrooms, Price of the apartment, parking availability (yes/no), direction facing, listed by (broker/individual), furnishing status (fully furnished/semi furnished/not furnished), facilities (gym, swimming pool, wifi, 24 hr electricity backup, etc.) and more information. In this paper, we have taken publicly available data manually from the website [9] (accessed on 25$^{th}$ March, 2019), which in turn collates the listings data from various publicly available sources. The total number of observations in the dataset is 66.

## 4. Experimental Results

Our goal is to solve the regression problem where the target variable is the price and the independent variables are number of bedrooms, number of bathrooms, super built up area, carpet area, furnishing status, floor type, direction facing, type of apartment, region and facilities. We perform one-hot encoding on facilities i.e. create categorical variables for each facilities. For example the feature vector for Apartment 1 with gym, swimming pool, wifi but no cafeteria would be [1 1 1 0] whereas the feature vector for Apartment 2 with gym, no swimming pool, no wifi and no cafeteria would be [1 0 0 0]. The number of independent variables is 56. We divide the data into training and test in the ration of 4:1. While building the model on the training dataset, we use 5 fold cross validation for all the methods. All the experiments were run using the open-source software R and on a system with configurations 4 GB RAM Mac iOS 1.6 GHz Intel Core i5.

We then apply advanced machine learning algorithms such as Random forests and Gradient boosting [15]. In Random forest we use ntrees=5000 and refer to it as RF for the rest of this paper. For Gradient boosting ntrees=10000 and learning rate= 0.01 and we will refer to this method as GBM for the rest of this paper. We have also used Artificial neural networks [16] with different hyper-parameters. We use one design of neural network with 2 hidden layers and 100 hidden nodes in each, which we will refer as ANN1 for the rest of this paper. In ANN2, we use 3 hidden layers with 100 hidden nodes in each and finally in ANN3, we use 3 hidden layers with 200 hidden nodes in each. The activation function used is "Rectifier". The evaluation metrics used for comparing the different methods are mean percentage error and median percentage error on the test dataset. The reason for taking median as it is considered as a robust measure that is not affected by outliers.

Table 1 shows the model results. We can clearly see that the RF i.e. Random forest method is the best performer in terms of prediction accuracy.

Table 1: Model Results on the Test dataset

| Method | Mean Percentage Error | Median Percentage Error |
|---|---|---|
| RF | 27.48% | 16.25% |
| GBM | 41.62% | 32.77% |
| ANN1 | 76.15% | 65.43% |
| ANN2 | 70.51% | 52.56% |
| ANN3 | 58.03% | 44.36% |

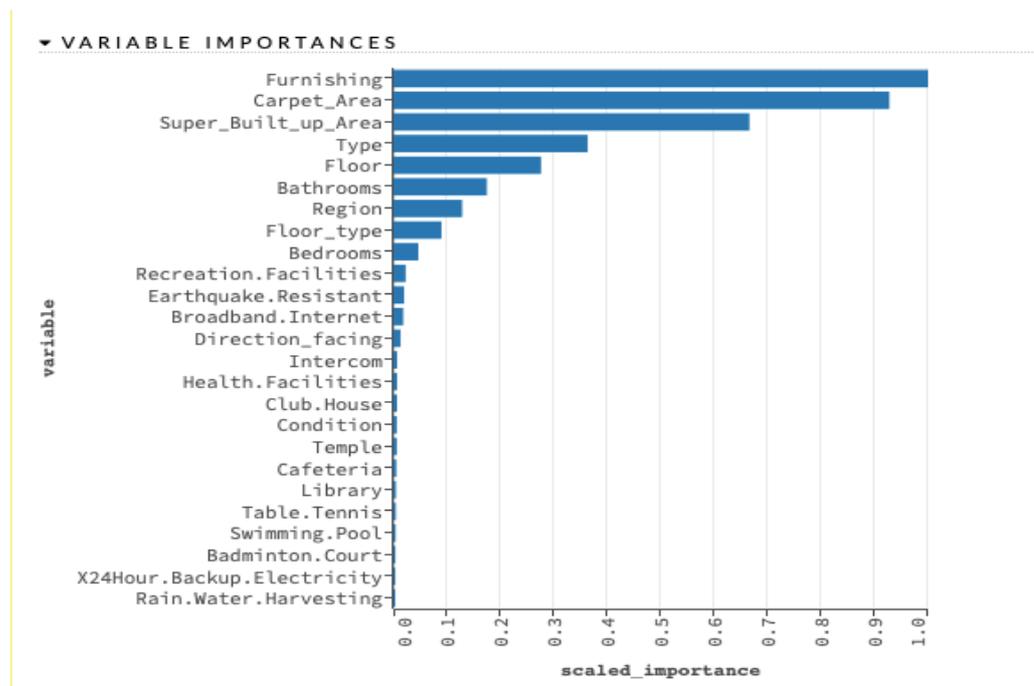

Figure 1: Variable Importance Chart

Figure 1 shows the variable importance plot for the RF method. We can see that the furnishing status, carpet area, super built up area, type of apartment, region, floor type, number of bathrooms and bedrooms are the most important variables.

## 5. Conclusion

In this paper, we have discussed the concept of PropTech and its applications. We worked on the use-case of proactive pricing of houses in classified advertisements in the Indian Real Estate market. We used advanced machine learning and artificial intelligence techniques to develop an algorithm that can predict housing prices based on certain input features. The business application of this algorithm is that classified websites can directly use this algorithm to predict prices of new properties that are going to be listed by taking some input variables and predicting the correct and justified price i.e. avoid taking price inputs from customers and thus keeping transparency in the system. The proposed methods can be used for other applications of PropTech as well as mentioned in Section 1. Since this was a proof of concept (POC), so the models were implemented on a smaller dataset. However, the error margins can be reduced further if we use much larger datasets, which we aim to work on in the future.

*This study is a proof-of-concept (PoC) and can be treated as a Technical Report. The purpose is to raise awareness about PropTech by solving one use-case.*

*\*This study is a proof-of-concept (PoC) and can be treated as a Technical Report. The purpose is to raise awareness about PropTech by solving one use-case.*